\documentclass[10pt,letterpaper,twocolumn]{revtex4-1} 

\usepackage{amsmath}
\usepackage{graphicx}
\begin{document}


\title{CaF$_2$ Whispering-Gallery-Mode Resonator Stabilized Narrow Linewidth Laser}


\author{B. Sprenger, H. G. L. Schwefel$^{*}$, Z. H. Lu, S. Svitlov, and L. J. Wang}

\address{
Max-Planck-Institute for the Science of Light\\
G\"unther-Scharowsky-Stra{\ss}e 1, Bau 24, 91058 Erlangen, Germany\\
$^*$Corresponding author: +49(9131) 6877-227, Harald.Schwefel@mpl.mpg.de
}

\begin{abstract}A fiber laser is stabilized using a Calcium Fluoride (CaF$_2$) whispering-gallery-mode resonator. It is set up using a semiconductor optical amplifier as a gain medium. The resonator is critically coupled through prisms, and used as a filtering element to suppress the laser linewidth. Using the self-heterodyne beat technique the linewidth is determined to be 13~kHz. This implies an enhancement factor of $10^3$ with respect to the passive cavity linewidth. The three-cornered hat method shows a stability of $10^{-11}$ after 10~$\mu$s.\end{abstract}


\maketitle

\noindent Whispering-gallery-mode (WGM) resonators have become popular in optics due to their compact sizes and their high quality (Q) factors\cite{vahala2003}. Typical resonators include molten microspheres\cite{braginsky1989}, on-chip disks and torroids\cite{little1997}, and crystalline disks\cite{grudinin2006}. The highest Q factors around $10^{10}$ have been demonstrated in CaF$_2$ crystalline disks.

Due to their high Q factors, WGM resonators can act as very precise filters\cite{matsko2007}. Sharp resonances can be filtered out by a resonator, or, as in this experiment, a narrow frequency selective element can be created by transmitting through a resonator. Before reaching the fundamental thermal sensitivity limit\cite{numata2004}, the stability of a reference cavity is mainly determined by its vibration.

\begin{equation}\label{eq:Young}
\frac{\Delta\nu}{\nu_0} = \frac{\Delta l}L = \frac{\rho \cdot \Delta g \cdot L}Y,
\end{equation}

\noindent where $\Delta \nu$ is the laser linewidth, $\nu_0$ is the central frequency, $\Delta l$ is the length change due to vibration, $L$ is the total length, $\rho$ is the density, $\Delta g$ is the vibration acceleration, and $Y$ is Young's modulus. From Eq.\ref{eq:Young} we can see that WGM resonators are excellent choices for frequency standards due to their small size.

Doped WGM resonators have been used in the past to lase directly, including rare-earth doped glass spheres\cite{miura1996}, and droplets with dye or quantum dots\cite{schafer2008}. In these techniques the Q factor is inherently limited due to absorption. Passive WGM resonators have been used in the past to stabilize lasers, although these often require additional frequency selective elements such as fiber Bragg gratings\cite{kieu2006,kieu2007}.

In this Letter, we report the observation and measurement of narrow-linewidth lasing from a semiconductor optical amplifier (SOA) fiber laser, stabilized using a CaF$_2$ resonator. The broad emission from the SOA is evanescently coupled into a CaF$_2$ disk using a prism, and out on the other side using another prism. Previous experiments have used tapered fibers, angle-polished fibers, and microspheres\cite{kieu2006,kieu2007,sprenger2009}. In this case CaF$_2$ disks were chosen due to their higher Q factor, and prism coupling was used due to the setup rigidity. In each roundtrip, only sharp modes can transmit through the resonator and thus achieve gain. The resulting narrow linewidth emission is studied using the three-cornered-hat measurement technique\cite{gray1974}, as well as a self-heterodyne measurement using a 45~km delay line\cite{okoshi1980}.

\begin{figure}[b]
\centering
\includegraphics[width=8.4cm]{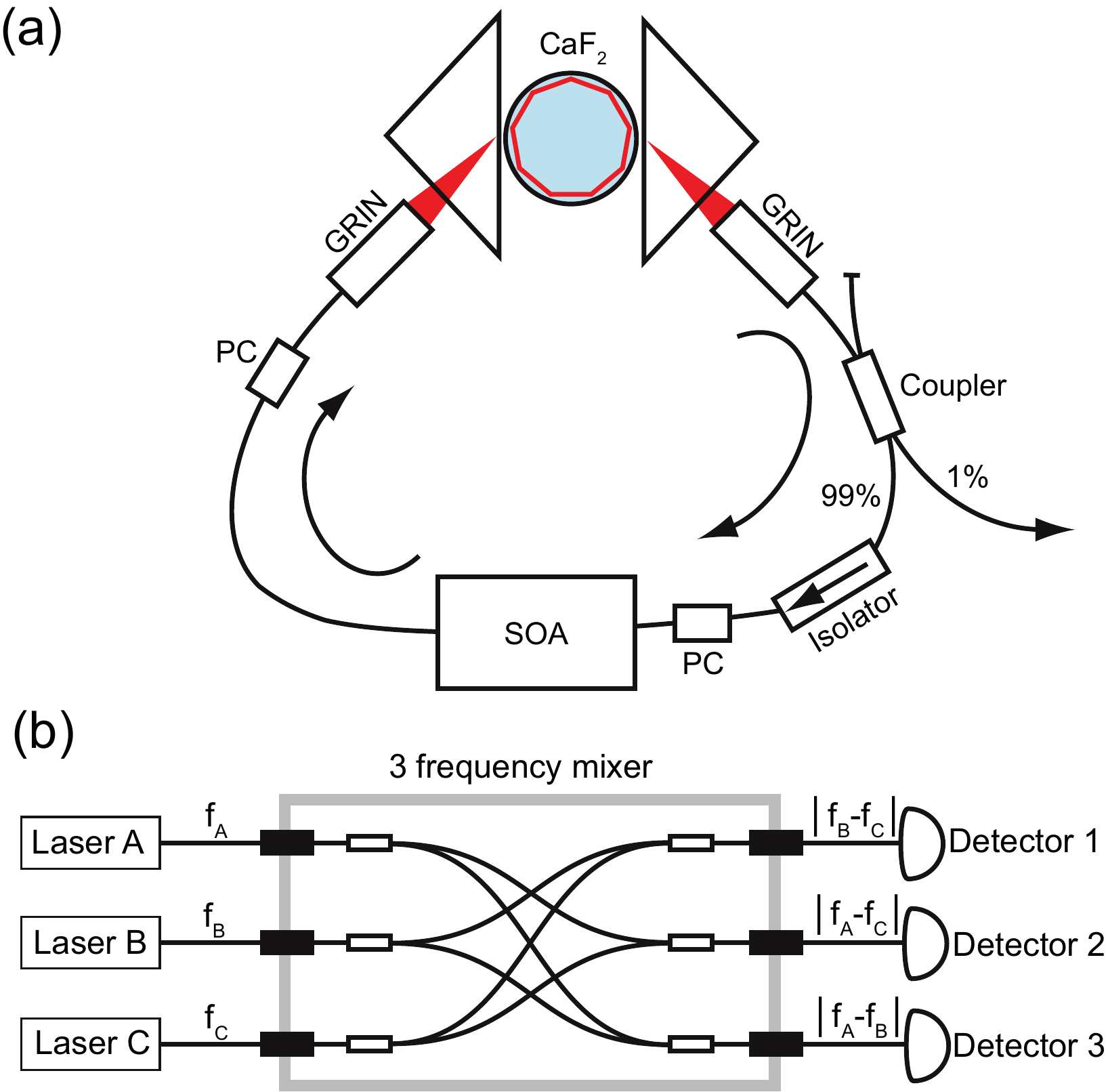}
\caption{(Color online) (a) Stabilized laser setup. Gain from a semiconductor optical amplifier (SOA) passes through polarization control (PC) and is coupled through a CaF$_2$ disc, using two prisms and graded refractive index (GRIN) lenses.  1\% of the 1550~nm emission is coupled out, and backwards lasing is prevented by an isolator. (b) Schematic of the frequency mixing for the three-cornered-hat measurement.}\label{fig:SetupFig}
\end{figure}

The CaF$_2$ resonator was turned and polished on a homemade machine to a diameter of about 5~mm. This results in a free spectral range of about 15~GHz at 1550~nm. The resulting passive linewidth is measured to be 15~MHz, which corresponds to a Q factor of $10^7$. Further investigations into higher Q factor resonators are planned. A graded refractive index (GRIN) lens focuses the spontaneous emission from the SOA (Thorlabs S9FC1004P) onto the back of a prism (SFL11 glass), which is at critical coupling distance from the disk. The same method is used on the other side of the prism to collect the transmitted mode, as shown in Fig.~\ref{fig:SetupFig}(a). The CaF$_2$ disk is mounted on a thermoelectric cooling element to control the temperature. Polarization control is used to optimize the polarization mode going into the resonator, as well as into the SOA for maximum gain. An isolator prevents a backwards propagating mode, and a fiber coupler emits 1\% of the lasing around 1560~nm. Single-mode lasing is achieved by carefully adjusting the coupling on each side of the resonator. Other gain materials which run single-mode in standard telecom fiber can be used in this setup. The SOA has an emission peak around 1560~nm, therefore a mode around this wavelength is selected automatically through mode competition in the laser. An Erbium-doped fiber pumped by a 980~nm laser can also be used instead of the SOA, resulting in single-mode lasing around 1530~nm without further re-adjustment.

\begin{figure}[tb]
\centering
\includegraphics[width=8.4cm]{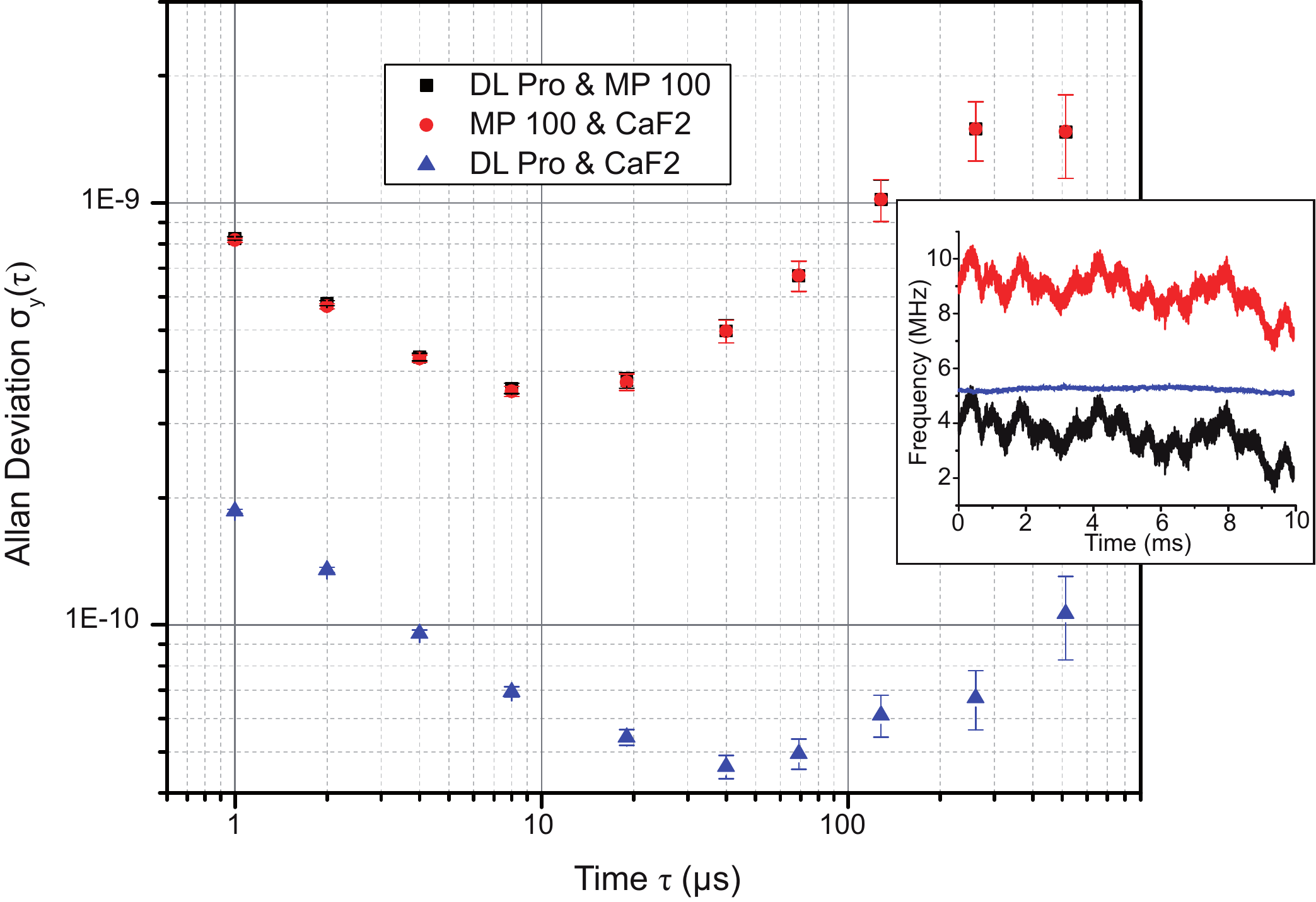}
\caption{(Color online) Allan deviations of simultaneous beat signals between a Toptica DL Pro, a home-built grating-stabilized diode laser (MP 100), and the CaF$_2$ disc stabilized laser. A common 1.5~kHz harmonic due to vibration is removed from the beats with the MP 100. Inset, frequencies of the beats as a function of time (10~ms). MP 100 and CaF$_2$, DL Pro and CaF$_2$, then DL Pro and MP 100 from top to bottom.}\label{fig:CombinedFig}
\end{figure}

Common methods to determine linewidths are heterodyne beating with a reference laser with similar stability, self-heterodyne beating using a delay line, and the three-cornered-hat method. The linewidth of the reference laser (Toptica DL Pro, 100~kHz) is larger than the linewidth of the CaF$_2$ stabilized laser, so the heterodyne beat technique is insufficient. The three-cornered-hat method has been used to determine absolute precision of frequency standards in the microwave domain such as Cesium or Hydrogen clocks\cite{gray1974}. More recently, experiments have been done in the optical regime to determine the absolute performance of narrow linewidth lasers\cite{lopez2006,zhao2009}. In this case, three lasers are used in the comparison: the CaF$_2$ disk stabilized laser, a Toptica DL Pro diode laser at 1550~nm, and a homebuilt grating stabilized diode laser in the Littrow configuration (MP 100). Outputs from the lasers are beat with each other as shown in Fig.~\ref{fig:SetupFig}(b). The beat notes around 5 to 10~MHz are recorded simultaneously on three detectors at a speed of 100~MSamples/s. Up to 10~ms of data are recorded and frequencies are calculated from 1~$\mu$s segments, as shown in the inset of Fig.~\ref{fig:CombinedFig}. The Allan deviation, a commonly quoted value for the stability of a frequency standard\cite{barnes1971}, is calculated for each beat. The Allan deviation of the MP 100 and CaF$_2$ stabilized laser beat notes, and of the DL Pro and MP 100 beat notes are around an order of magnitude worse that of the DL Pro and the CaF$_2$ laser beat notes. This implies that the MP 100 has a larger linewidth than that of the DL Pro and the CaF$_2$ laser.

The Allan variance of a beat note is given by $\sigma_{ab}^2=\sigma_a^2+\sigma_b^2$, where $\sigma_a^2$ and $\sigma_b^2$ are the Allan variations of lasers a and b respectively. The same is true for the other two combined beats. By rearranging these formulas one finds the following to be true for the Allan variance of an individual frequency standard\cite{gray1974}.

\begin{equation}\label{Allan}
\sigma_a^2 = \frac 1 2 \cdot (\sigma_{ab}^2+\sigma_{ac}^2-\sigma_{bc}^2).
\end{equation}

\begin{figure}[tb]
\centering
\includegraphics[width=8.4cm]{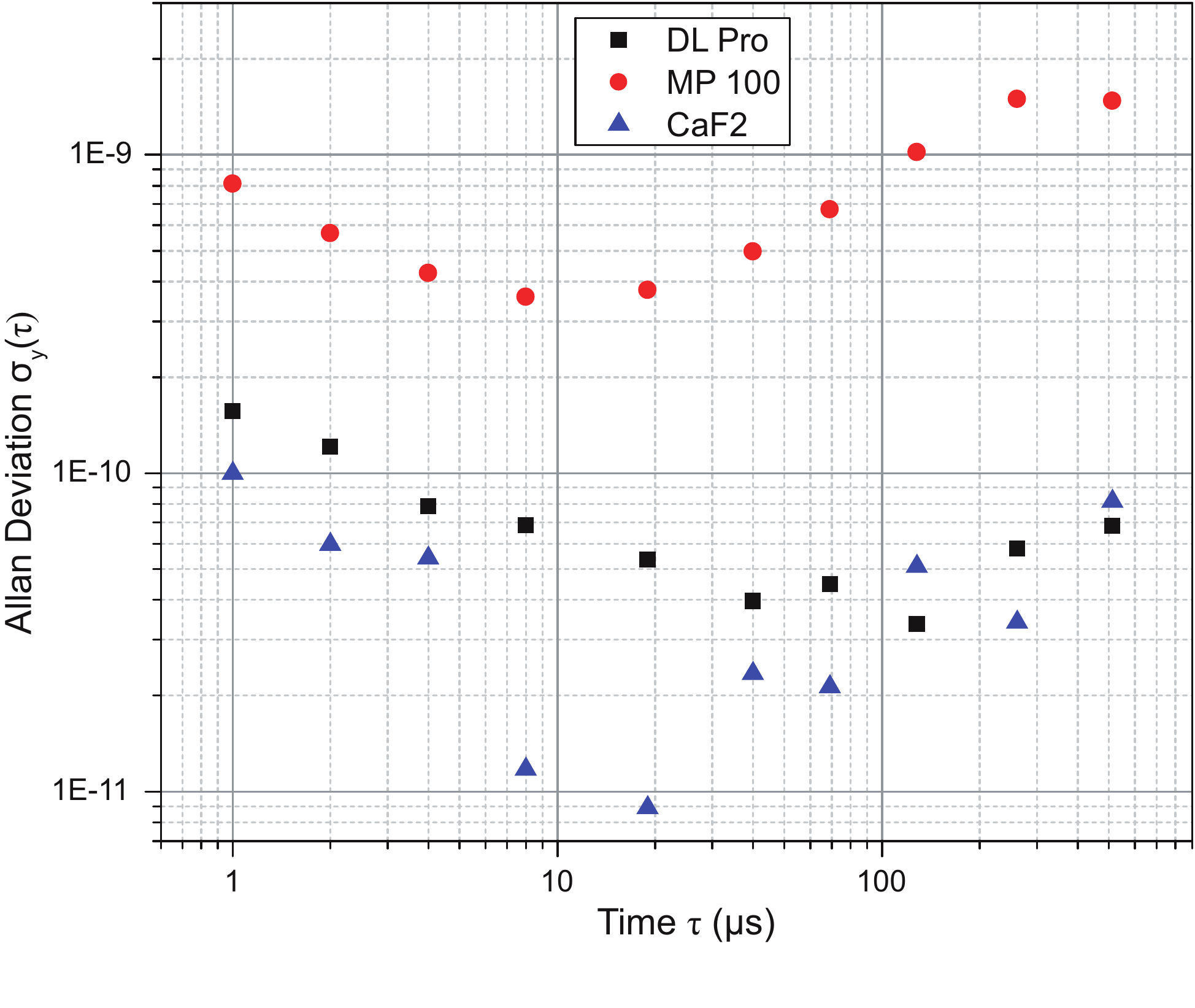}
\caption{(Color online) Allan deviations of the three individual lasers, calculated using the three-cornered-hat method.}\label{fig:ThreeCornerFig}
\end{figure}

Similar equations can be found for $\sigma_b^2$ and $\sigma_c^2$. The resulting Allan deviations for the lasers can be seen in Fig.~\ref{fig:ThreeCornerFig}. The CaF$_2$ resonator initially has the best performance, and is comparable to the performance of the DL Pro at longer averaging times. The MP 100 is clearly the worst of the three lasers. The stability of the CaF$_2$ disk stabilized laser reaches approximately $10^{-11}$ after 10~$\mu$s, implying a fast linewidth of 2~kHz.

\begin{figure}[tb]
\centering
\includegraphics[width=8.4cm]{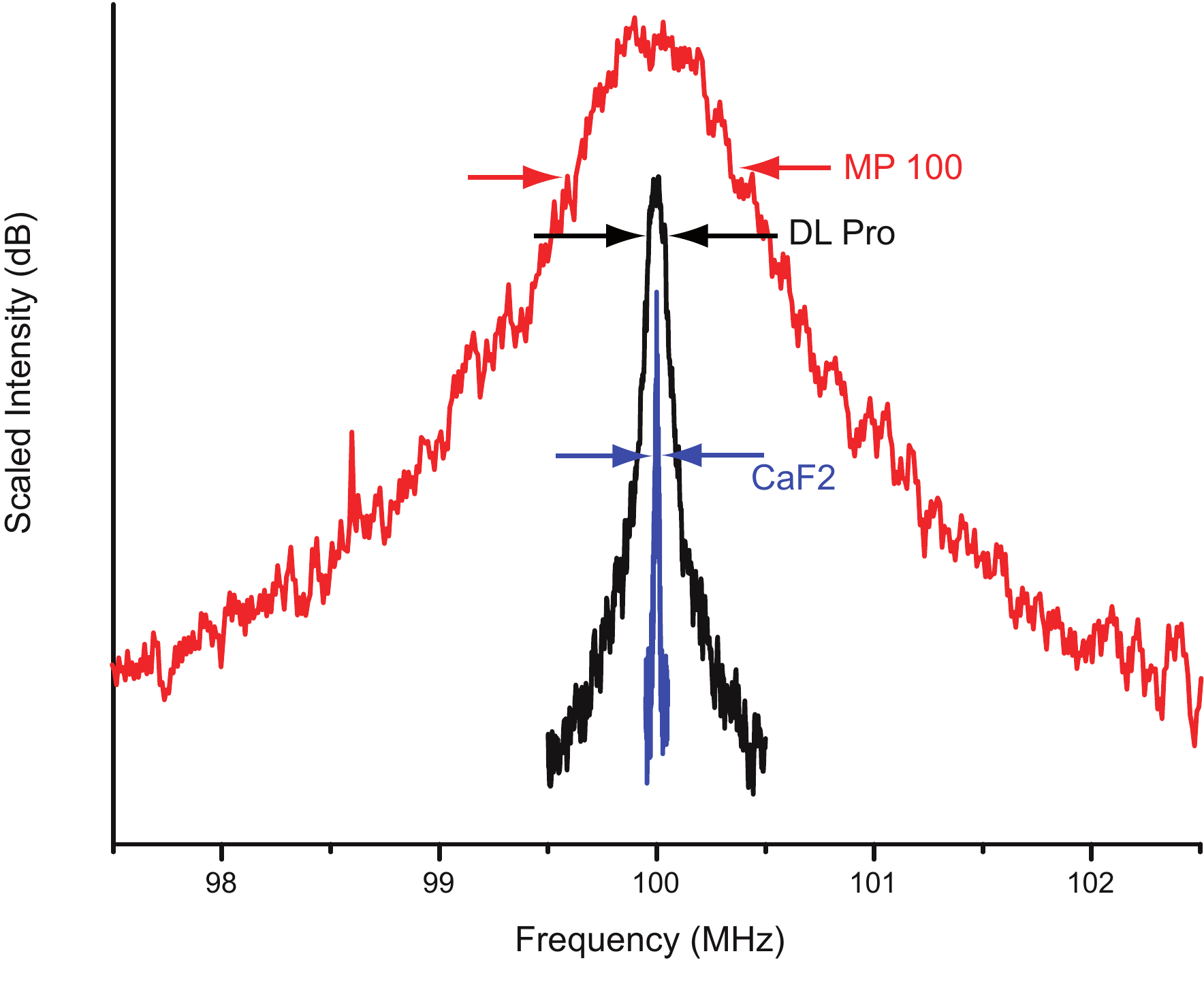}
\caption{(Color online) Self-heterodyne technique measurements. Beats are at the AOM frequency of 100~MHz. Recording spans of the MP 100, DL Pro, and CaF$_2$ stabilized laser are 5~MHz, 1~MHz, and 100~kHz respectively. 3~dB linewidths are 550~kHz, and 60~kHz, and 18~kHz. The delay is 45~km. Amplitudes are scaled for visibility, and the 3~dB widths are indicated by arrows.}\label{fig:HetFig}
\end{figure}

In order to verify these results, the self-heterodyne technique is employed. It is commonly used to determine narrow linewidths by splitting the laser signal using an acousto-optic modulator (AOM). One part, in this case offset by 100~MHz by the AOM, is sent through a 45~km delay line of fiber. Then it is recombined with the other part in a fiber coupler and the signal is detected on a fast photodiode (MenloSystems FPD510) and recorded using a spectrum analyzer. The delay length should be at least six times longer than the coherence length of the laser for a valid uncorrelated beat\cite{richter1986}. In a 10~kHz laser this corresponds to 36~km of fiber. To be sure 45~km of fiber are used. The signal to noise ratio is quite low, so 70 averages are used in the final result. For the CaF$_2$ resonator the span is 100~kHz, the resolution bandwidth is 1~kHz, and the sweep time is 100~ms. The span is set to 1~MHz for the DL Pro, and 5~MHz for the even wider grating stabilized diode laser. The results are shown in Fig.~\ref{fig:HetFig}. The 3~dB width of the grating stabilized laser is 550~kHz, and 60~kHz for the DL Pro. The WGM resonator stabilized laser has a self-heterodyne linewidth of 18~kHz. To calculate the separate linewidth one needs to divide by $\sqrt{2}$ for a Gaussian spectrum, and $2$ for a Lorentzian spectrum\cite{richter1986}. This implies that the upper limits for the laser linewidths are 390~kHz for the MP 100, 43~kHz for the DL Pro, and 13~kHz for the WGM stabilized laser. This corresponds to an enhancement factor of about $10^3$ from the original resonance width of 15~MHz for the CaF$_2$ disk. The result is an order of magnitude better than our previous linewidth\cite{sprenger2009}.

In conclusion, a passive fiber laser stabilization setup was presented. The high Q modes of a WGM resonator act as a filter, thus only allowing specific modes to achieve gain. Single-mode lasing can be induced using a gain material with a peak around 1550~nm, without the need of an additional frequency selective element. The linewidth was measured to be around 13~kHz using a CaF$_2$ resonator with a Q factor of just $10^7$. This implies an enhancement factor of over $10^3$, and is an order of magnitude better than previous experiments have shown. WGM resonators are an excellent candidate with their compact sizes and high Q factors for passive laser stabilization, as well as frequency references and metrology.

The authors would like to thank Y. N. Zhao, S. Preu, V. Elman, and J. Zhang for stimulating discussions.


\end{document}